\documentstyle[11pt,newpasp,twoside,epsf]{article}
\def\Sec{\hbox{${}^{\prime\prime}$\llap{.}}}

\def\sec{\hbox{${}^{\prime\prime}$}}
\def\kms{km s$^{-1}$}

\def\mh{{M_{\bullet}}}

\def\msun{{M$_{\odot}$}}

\def\ms{{\mh-\sigma}}

\def\ml{{\mh-L_B}}
\def\lae{\mathrel{<\kern-1.0em\lower0.9ex\hbox{$\sim$}}}
\def\gae{\mathrel{>\kern-1.0em\lower0.9ex\hbox{$\sim$}}}

\def\edcomment#1{\iffalse\marginpar{\raggedright\sl#1\/}\else\relax\fi}

\reversemarginpar

\begin{document}

\title{Supermassive Black Hole Research in the Post-HST Era}
\author{Laura Ferrarese}
\affil{Rutgers University, 136 Frelinghuysen Road, Piscataway, NJ 
08854}

\begin{abstract}

Thanks to its unprecedented spatial resolution, the Hubble Space
Telescope has ended a 20-year long stalemate by detecting the
dynamical signature of nuclear supermassive black holes (SBHs) in a
sizeable number of nearby galaxies. These detections have revealed the
existence of a symbiotic relationship between SBHs and their hosts,
changing the way we view SBH and galaxy formation.  In this
contribution I review which are the most pressing outstanding issues 
in SBH research, and what are the technological requirements needed to
address them.

\end{abstract}

\section{The Current Status of SBH Searches}

The study of supermassive black holes is one of the areas of modern 
astrophysics which has benefited most from the launch of HST. After 
two decades of tantalizing but inconclusive ground-based studies, the 
HST/FOS observations of M87 (Harms et al.  1994) and NGC 4261 
(Ferrarese et al.  1996) provided the first firm measurements of SBH 
masses in galactic nuclei.  In the years that followed, FOS and STIS 
data lead to detections in ten additional galaxies (Bower et al.  
1998; van der Marel \& van den Bosch 1998; Ferrarese \& Ford 1999; 
Emsellem et al.  1999; Cretton \& van den Bosh 1999; Verdoes Kleijn et 
al.  2000; Gebhardt et al.  2000a; Joseph et al.  2001; Barth et al.  
2001; Sarzi et al.  2001).  The superiority of HST over ground based 
facilities in this field is easily understood.  Only dynamical 
evidence, either from gas or stellar kinematics, can yield compelling 
proof of the existence of SBHs.  With rare exceptions (e.g.  M31, 
M87), ground based telescopes lack the spatial resolution necessary to 
resolve the SBH ``sphere of influence'', i.e.  the region of space 
within which the SBH gravitational influence dominates that of the 
surrounding stars:

\begin{equation}r_h = G\mh/\sigma^2 \sim 11.2(\mh/10^8~{\rm 
M_{\odot}}) / 
(\sigma /200~ {\rm km ~s^{-1}})^2 {\rm ~pc},\end{equation}

\noindent with $\sigma$ the stellar velocity dispersion and $\mh$ the 
SBH mass.  Resolving $r_h$ is a necessary condition for a SBH 
detection to be made; not meeting it leads to spurious detections and 
biased masses (Merritt \& Ferrarese 2001a).

With over a dozen secure measurements, it has become possible to 
search for correlations between $\mh$ and the overall properties of 
the host galaxies.  The first relation to emerge was one between $\mh$ 
and the blue luminosity $L_B$ of the surrounding bulge (Kormendy \& 
Richstone 1995).  A much tighter correlation was subsequently 
discovered between $\mh$ and the bulge stellar velocity dispersion 
(Ferrarese \& Merritt 2000; Gebhardt et al.  2000b):

\begin{equation}\mh = \beta {\left({\sigma} \over {200 {\rm 
~km~s^{-1}}}\right)}^\alpha.\end{equation}

\noindent with $\alpha = 4.58 \pm 0.52$ and $\beta = (1.66 \pm 0.32) 
\times 10^8$ \msun~ (Ferrarese 2002a).  More recently, evidence has 
emerged that a fundamental relation might exist between $\mh$ and the 
mass $M_{DM}$ of the dark matter halos in which the SBHs presumably 
formed (Ferrarese 2002b):

\begin{equation}{{\mh} \over {10^8~{\rm M_{\odot}}}} \sim 0.10 
{\left({M_{DM}} 
\over {10^{12}~{\rm  {\rm M_{\odot}}}}\right)}^{1.65}\end{equation}

The above relations have proven invaluable in the study of SBH 
demographics (Merritt \& Ferrarese 2001b; Ferrarese 2002a; Yu \& 
Tremaine 2002) and have generated intense activity on the theoretical 
front (Haehnelt, Natarajan \& Rees 1998; Silk \& Rees 1998; Cattaneo, 
Haehnelt \& Rees 1999; Adams, Graff \& Richstone 2000; Monaco et al.  
2000; Haehnelt \& Kauffmann 2000; Wyithe \& Loeb 2002).  At the same 
time, new questions have arisen, and with them the need for further 
observational constraints.  In this contribution, I will address three 
such questions, and identify the technological requirements necessary 
to answer them.

\begin{itemize}

\item {\it What are the characteristics of the $\ms$ relation?  Does the 
slope and/or normalization of the relation depend on Hubble type, 
environment, and/or redshift?} As theoretical models are refined, 
tighter observational constraints will be required.  Most of the SBHs 
detected to date are in the $10^8 \lae \mh \lae 10^9$ range.  The 
$\ms$ relation is not sampled below $10^6$ \msun, and badly sampled 
for $10^6 \lae \mh \lae 10^7$ \msun.  There are few spiral galaxies 
represented, all of which are early type, and only two galaxies well 
beyond 30 Mpc.

\item {\it Are binary supermassive black holes long lived?} The 
existence and lifespan of binary SBHs can have dramatic consequences, 
from shaping the morphology and dynamics of the resulting galaxy 
(Milosavljevic \& Merritt 2001; Milosavljevic et al.  2002; 
Ravindranath et al.  2002; Yu 2002) to destroying nuclear dark matter 
halo cusps (Merritt et al.  2002).

\item {\it How small can nuclear BHs be?  Are there nuclear BHs in 
globular clusters?} There is no dynamical evidence for ``intermediate 
mass'' black holes (IBHs) in the $\mh \sim 10^2 - 10^6$ \msun~range, 
although their existence in the off-nuclear regions of some starburst 
galaxies is supported by energetic arguments (Fabbiano et al.  2001; 
Matsumoto et al.  2001).  There is also no dynamical evidence that BHs 
are formed in the nuclei of globular clusters (van der Marel et al.  
2000; Gebhardt et al.  2000).  However, whether such black holes exist 
is critical for our understanding of how SBHs form.  In ``top-down'' 
self-regulating models that trace the formation of SBHs to the very 
early stages of galaxy formation, there is a natural lower limit of 
$\sim 10^6$ \msun~to $\mh$ (e.g.  Loeb 1993; Silk \& Rees 1998; 
Haehnelt, Natarajan \& Rees 1998).  On the other hand, in 
``bottom-up'' models nuclear SBHs are formed by the merging of IBHs.  
These are deposited at the galactic center as the globular clusters in 
which they originally formed spiral in due to dynamical friction 
(Portegies Zwart \& McMillan 2002; Ebisuzaki et al.  2001).  In the 
latter scenario, no physical reason would prevent the formation of 
SBHs with $\mh \lae 10^6$ \msun.

\end{itemize}

\section{Building the Local Sample}

\begin{figure}[t]
\plotfiddle{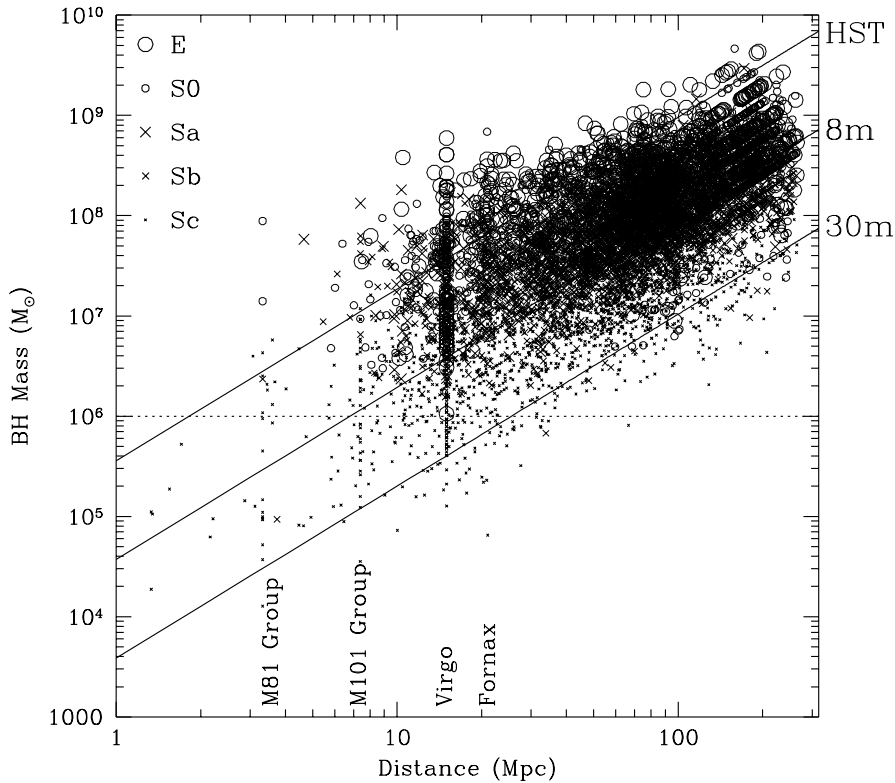}{4in}{0}{63}{63}{-190pt}{-110pt}
\caption{SBH mass vs. distance for all galaxies in the CfA Redshift
Sample (Huchra et al. 1990). Only for the galaxies which lie above the
solid lines the sphere of influence of the putative nuclear SBH can be
resolved by HST/STIS, an 8m and a 30m diffraction limited telescope.
A few nearby groups and clusters are marked. It should be noted that 
because of the large scatter in the $\ml$ relation (a factor of 
several in $\mh$), this figure has only statistical value.  }
\end{figure}

Galaxies come in all flavors, but do all flavors come with a SBH?  The
sample of galaxies within the reach of HST is remarkably
homogeneous. Most are early type galaxies. And although we might
expect the number of detections to double or perhaps even triple as
new HST data become available (Kormendy \& Gebhardt 2001), most will
remain in the $10^8 \lae \mh \lae 10^9$ \msun~ range which is already
well sampled by the current data (Merritt \& Ferrarese 2002a). Hopes
of breaking the $10^6$ \msun~barrier lie in one galaxy only, NGC 205,
a compact elliptical already scheduled to be observed with HST
(P.I. L. Ferrarese)\footnotemark.

\begin{figure}[t]
\plotfiddle{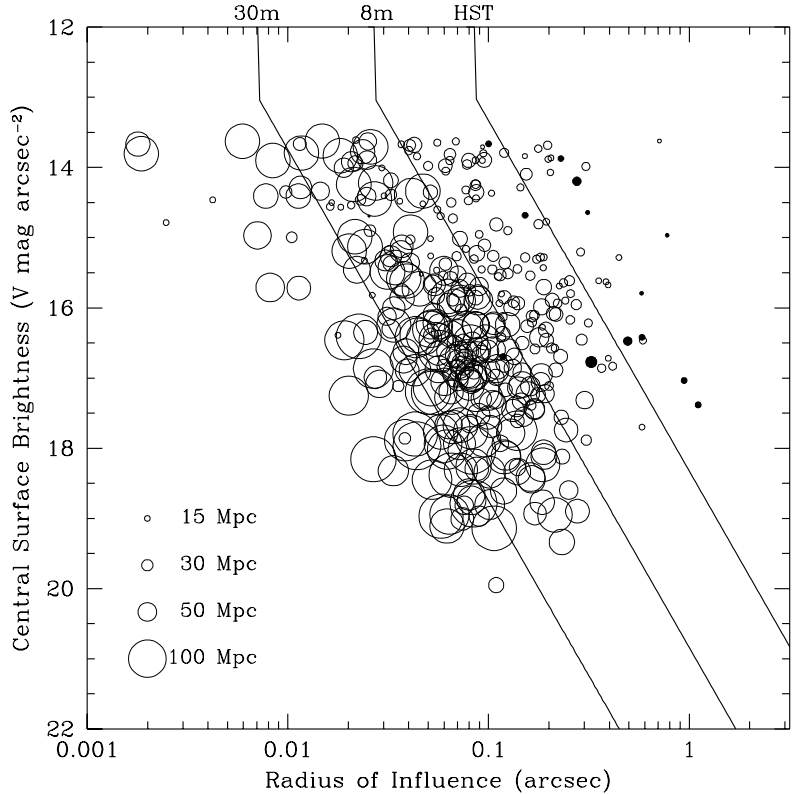}{4in}{0}{63}{63}{-190pt}{-110pt}
\caption{SBH radius of influence vs. central surface brightness for
the Early Type galaxy sample of Faber et al. (1989). The size of the
symbols is proportional to the  galaxy distances, as shown in the
legend. Spectra with ideal signal-to-noise and spatial resolution for
dynamical studies can be collected in the equivalent of 3 HST orbits
only for the galaxies to the right of the solid lines (shown for
HST/STIS, an 8m and a 30m diffraction limited telescopes). The solid
circles identify galaxies observed with HST.}
\end{figure}

\footnotetext{Based on the $\ms$ relation, NGC 205 is expected to host
a $\sim 7.5\times 10^4$ \msun~black hole; at a distance of 740 kpc, a
black hole as small as $6 \times 10^4$ \msun~can be detected by STIS.}

But there are other barriers HST cannot break.  The low central 
surface brightness which characterizes giant ellipticals (e.g.  
Ferrarese et al.  1994; Lauer et al.  1995; Rest et al.  2000) makes 
stellar dynamical studies with HST prohibitive.  For instance, 
measuring $\mh$ in M87 ($d \sim 15$ Mpc) using stellar dynamics would 
require over 100 orbits of STIS time\footnotemark.  While this problem 
can be avoided by using gas, rather than stellar, kinematics, only a 
fraction of ellipticals host the regular NGC4261-like dust disks 
(Jaffe et al.  1993) which make gas dynamical studies possible.  
Therefore, ironically, most of the very largest SBHs ($\mh \gae 10^9$ 
\msun) are beyond the reach of HST.

\footnotetext{The calculation assumes S/N=50 at $\sim 8500$ \AA.}

Similarly, the vast majority of dwarf elliptical galaxies are beyond 
HST capabilities.  Assuming that the $\ms$ relation holds in the 
$\sigma \sim$ a few $\times 10$ \kms~regime that characterizes these 
objects, the sphere of influence of the putative SBH at their centers 
is accessible only in the most nearby systems.  In these cases, the 
stellar population is resolved into individual stars, each too faint 
to be handled by HST's small mirror.  For instance, a constraint on 
the central mass in NGC 147 ($d \sim 660$ kpc) would require several 
hundred orbits with STIS. The situation for late type spirals is no 
better.  The upper limit on $\mh$ in M33, the closest Sc galaxy, puts 
the sphere of influence of the putative BH a factor $\sim 20$ below 
the resolution capabilities of HST.

The above statements are quantified in Figure 1.  For each galaxy in 
the CfA redshift sample (Huchra et al.  1990), $\mh$ is calculated 
from the $\ml$ relation given by Ferrarese \& Merritt (2000), after a 
correction suitable to each Hubble type is applied to convert total 
luminosity to bulge luminosity (Fukugita et al.  1998).  If only 
resolution constraints are considered, most late type spirals (Sb - 
Sc) are expected to host SBHs too small to be resolved by HST. Only a 
handful of SAs are within reach.  It is only with an 8m class 
telescope that a complete sample of galaxies spanning the whole Hubble 
sequence can be collected.  Even then, little would be gained below 
$10^6$ \msun: pushing this limit down by an order of magnitude 
requires a 30m diffraction limited telescope.

The above constraints become tighter when exposure time requirements
are folded in. Figure 2 shows the detection limits for HST, an 8m and
a 30m diffraction limited telescopes assuming the equivalent of 3 HST
orbits of integration on each galaxy$^2$.  The points represent the
early type galaxy sample of Faber et al. (1989), with $\mh$ calculated
using the $\ms$ relation. Again, HST can only see the top of the
iceberg, producing enough SBH detections to allow us a glimpse of what
lays underneath, but leaving most of the parameter space
unexplored. To study the influence of environment on the formation and
evolution of SBHs, it would be helpful to probe  Coma and the most
nearby rich Abell clusters at $d \sim 100$ Mpc.  In order to push
farther and perhaps study the redshift evolution of the $\ms$
relation, reverberation mapping (see Brad Peterson's contribution in
these proceedings) will likely remain the only viable method.

\section{Binary Supermassive Black Holes}

\begin{figure}[t]
\plotfiddle{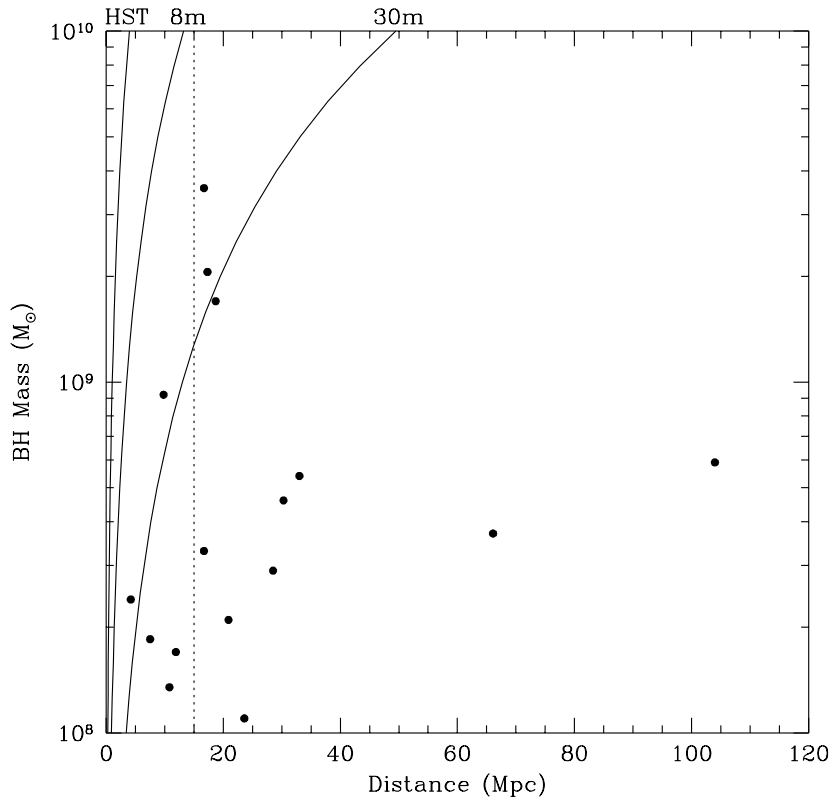}{4in}{0}{63}{63}{-190pt}{-110pt}
\caption{The points represent all SBH masses detected to date (see
references in Merritt \& Ferrarese 2000). According to the simulations
of Milosavljevic \& Merritt 2001, a SBH would have to lie to the left
of the solid line (drawn for HST, an 8m and a 30m diffraction limited
telescope) to be resolved as a binary. The vertical dotted line marks
the position of the Virgo cluster.}
\end{figure}

The formation of binary SBHs as a consequence of merging seems 
unavoidable (Begelman, Blandford \& Rees 1980), however the evolution 
of the SBH binary after merging is very uncertain (Ebisuzaki, Makino 
\& Okumura 1991; Milosavljevic \& Merritt 2002; Yu 2002).  The recent 
work of Milosavljevic \& Merritt (2002) represents the state of the 
art in numerical simulations of merging galaxies with SBHs.  We will 
adopt the results of this work to estimate the resolution requirements 
necessary to detect the dynamical signature of a SBH binary.  
Milosavljevic \& Merritt follow the merging of two low-luminosity 
ellipticals, characterized by steep power-law stellar density 
profiles.  Well within one Myr after the galaxies merge, the two SBHs 
form a hard binary with separation between a few hundredths to a few 
parsecs, depending on the SBHs masses.  By exchanging energy with 
nearby stars, the binary will start to harden; however, Milosavljevic 
\& Merritt find that the hardening will eventually stall, as a 
consequence of the depletion of stars with which the binary can 
interact.  At this stage, the separation between the SBHs is $\lae 
0.2$ pc ($\propto \mh^{0.57}$), and the stellar rotational velocity 
and velocity dispersion differ significantly from the single SBH case 
(see Fig.  15 of Milosavljevic \& Merritt).  Figure 3 shows the limit 
at which the ``final'' distance between the binary SBHs can be 
resolved with HST, an 8m and a 30m diffraction limited telescope, as a 
function of the binary mass and distance.  The solid points show all 
SBHs detected to date.  Rather obviously, a direct dynamical detection 
of a binary SBH would require a 30m class telescope, and even then 
only a few of the most nearby galaxies would be accessible.  Unless 
proof of the existence of binary SBHs can be obtained in some other 
way (with LISA, for instance), it would not be a bad idea to start 
cheering for OWL (see the contribution by Roberto Gilmozzi in these 
proceedings).

\section{Black Holes in Globular Clusters}

The formation of BHs at the centers of highly concentrated globular 
clusters (GCs) is suggested by theoretical arguments and numerical 
simulations (Miller \& Hamilton 2001; Mouri \& Taniguchi 2002; 
Portegies Zwart \& McMillan 2002).  Detecting the signature of a BH in 
GCs is best done through proper motion, rather 
than spectroscopic studies.  The latter have been pursued in the case 
of M15, with no conclusive results (van der Marel 2000; Gebhardt et 
al.  2000).  Velocities of several tens to several hundred stars 
within the sphere of influence of the putative BH must be collected in 
order to derive an accurate value of the velocity dispersion; 
unfortunately at the Mg b triplet ($\sim 5500$ \AA), and even more 
severely at the Ca II triplet ($\sim 8500$ \AA), the fainter turnoff 
and main sequence stars are drowned in the light of a few nearby 
giants.  In the case of M15, for instance, the sphere of influence of 
a $10^3$~\msun~BH is $\sim 1$\sec, and about 90 velocities are needed 
to produce a value of the velocity dispersion accurate to 10\% within
this region; using state of the art ground-based instrumentation, 
Gebhardt et al.  (2000) were able to measure velocities for only 5 
stars within 1\sec.

The suitability of proper motion studies in constraining the central 
potential has been demonstrated in spectacular fashion for the 
Galactic center (Ghez et al.  2000; Genzel et al.  2000), using ground 
based facilities.  With HST, the art of extracting astrometric 
information from WFPC2 images has been perfected by Anderson \& King 
(2000), who can reach an astonishing 0.005 pixel positional accuracy 
using properly dithered data.  Constraints on the BH mass at the 
center of Galactic globular clusters can be easily calculated; under 
appropriate assumptions, with two ten-orbit exposures, taken one year 
apart, HST could set limits of 2000 \msun~ or better on the BH masses 
at the center of $\sim$ ten of the nearest Galactic GCs.  For a few 
GCs, the limit is below 1000 \msun, well in the range of BH masses 
estimated for the off-nuclear IBHs in the Antennae galaxies and M82 
(Fabbiano et al.  2001; Matsumoto et al.  2001).  While even tighter 
constraints could be reached using a larger aperture, the combination 
of HST and ACS can be very competitive.

\begin{table}
\caption{Technical Requirements}
\begin{tabular}{llllll}
\tableline
\tableline
\multicolumn{6}{l}{\bf Project: Enlarging the Sample and Probing $< 
10^6$ and $> 10^9$ \msun~ SBHs}\\
\multicolumn{6}{l}{\it Needed to study systematics, distinguish between
``bottom-up'' and ``top-down'' models of}\\ 
\multicolumn{6}{l}{\it  SBH formation, constrain the role of feedback in SBH accretion during merging}\\
Resolution & Aperture & FOV & $\lambda$ range & $\lambda/\Delta\lambda$ & Comments \\
$<$0\Sec02 & $>$ 8m & few$\times 10\sec$ & 5500$-$9500 \AA & 10,000 & Longslit\\
& & & & & IFU highly desirable\\
\tableline
\multicolumn{6}{l}{\bf Project: Resolving SBH Binaries}\\
\multicolumn{6}{l}{\it Needed to contrain dynamical models of galaxy mergers, determine 
the impact of SBH}\\
\multicolumn{6}{l}{\it  binaries in the morphological evolution of galaxies, contrain accretion mechanism.}\\
Resolution & Aperture & FOV & $\lambda$ range & $\lambda/\Delta\lambda$ & Comments \\
$<$0\Sec007 & $>$ 30m & few$\times 10\sec$ & $> 8000$ \AA & 10,000 & IFU\\
\tableline
\multicolumn{6}{l}{\bf Project: Detecting BHs in GCs}\\
\multicolumn{6}{l}{\it Needed to constrain dynamical models for SBH evolution, investigate
 the connection}\\
\multicolumn{6}{l}{\it  between GCs, nuclear SBHs, and galactic bulges.}\\
Resolution & Aperture & FOV & $\lambda$ range & $\lambda/\Delta\lambda$ & Comments \\
$<$0\Sec05 & $>$ 2.5m & few$\times 10\sec$ & $U$ or $B-$band  & N/A & High Resolution Imager,\\
& & & & & High Dynamic Range\\
\tableline
\tableline
\end{tabular}
\end{table}

Table 1 summarized the observational requirements needed to answer the
questions posed in the introduction of this contribution.  While most
applications will require higher spatial resolution and a larger
collective area, the detection of BHs in GCs could indeed be the next
legacy of HST to SBH research.

\end{document}